# Characterizing soot in TEM images using a convolutional neural network

Timothy A. Sipkens[*,1,2], Max Frei[3,4], Alberto Baldelli[1,5], P. Kirchen[1], Frank E. Kruis[3,4], Steven N. Rogak[1]

[1] Department of Mechanical Engineering, University of British Columbia, Vancouver, BC, Canada
[2] Department of Mechanical Engineering, University of Alberta, Edmonton, AB, Canada
[3] Institute of Technology for Nanostructures, University of Duisburg-Essen, Duisburg, NRW, Germany
[4] Center for Nanointegration Duisburg-Essen, University of Duisburg-Essen, Duisburg, NRW, Germany
[5] Faculty of Land and Food Systems, University of British Columbia, 2357 Main Mall, Vancouver, BC, Canada V6T 1Z4
* Corresponding author, tsipkens@uwaterloo.ca

**Abstract**

Soot is an important material with impacts that depend on particle morphology. Transmission electron microscopy (TEM) represents one of the most direct routes to qualitatively assess particle characteristics. However, producing quantitative information requires robust image processing tools, which is complicated by the low image contrast and complex aggregated morphologies characteristic of soot. The current work presents a new convolutional neural network explicitly trained to characterize soot, using pre-classified images of particles from a natural gas engine; a laboratory gas flare; and a marine engine. The results are compared against other existing classifiers before considering the effect that the classifiers have on automated primary particle size methods. Estimates of the overall uncertainties between fully automated approaches of aggregate characterization range from 25% in $d_{p,100}$ to 85% in $D_{TEM}$. A consistent correlation is observed between projected-area equivalent diameter and primary particle size across all of the techniques.

## 1. Introduction

Soot, a class of carbonaceous particles often produced by combustion processes, is one of the largest uncertainties to climate change models [1, 2] and has significant impacts on human health [3]. The particles themselves typically form fractal aggregate structures, which complicate characterization and modeling (e.g., of their optical properties [4, 5]). Within this context, transmission electron microscopy (TEM) is a critical tool, providing one of the most direct routes to examining particle morphology [6]. However, quantitative analysis of the particle population requires robust image analysis, and statistically significant results require a substantial sample of particles, which is often a labour-intensive task that requires substantial user intervention. While this has led to a range of automated or semi-automated methods aimed at reducing the cost associated with image analysis and enabling characterization of higher numbers of particles, the lower contrast of carbonaceous particles against the carbon films on which they are often captured combined with the relatively complex morphology of the particles has hindered attempts at fully automating the procedure. It also means that previous analysis tools developed for engineered nanoparticles (e.g., Pebbles [7] or the approaches in Refs. [8, 9]) are likely to be ineffective.

The analysis procedure to characterize aggregates can be broken into two steps (e.g., [10]): (*i*) semantic segmentation at the aggregate level, where the image is separated into background and aggregate pixels, and (*ii*) determining the primary particle size or other morphological parameters. Aggregate-level segmentation is generally the easier of the two tasks, with various methods available. The output is a binary mask that has immediate utility in computing the projected area of the aggregates, a useful parameter in gauging the particle size distribution. The binary mask is also often a direct input to other techniques aimed at computing the primary particle size – including the pair correlation method (PCM) [11]; Euclidean distance mapping, surface-based scale analysis (EDM-SBS) [12]; the hybrid Euclidean distance mapping-watershed analysis (EDM-WS) of De Temmerman et al. [13]; and the Hough transform method of Grishin et al. [14] – and other morphological parameters of each aggregate. Even when not explicitly required as an input – including for the Hough transform method proposed by Kook et al. [15] or Altenhoff et al. [16], the relative optical density method of Tian et al. [17], or the center-selected edge scoring method (CRES) of Andersen et al. [18] – subsequent processing steps can use the binary masks to identify which features in the image belong to which aggregate. This, in turn, enables study of size-resolved characteristics, including trends in the effective density [19, 20] that can be compared to other population-based diagnostics, e.g., [21, 22, 23].

Various approaches have been proposed in an attempt to automatically segment the aggregates from the background, hereafter referred to as *classifiers* (in contrast to the methods used for primary particle sizing techniques). The simplest is thresholding, where all of the pixels with an intensity below a certain value are considered part of the aggregate. There are numerous automated methods to estimate an optimal threshold in the image analysis realm (cf. Sezgin and Sankur [24]). The most prominent of these automated methods is arguably Otsu's thresholding [25], which has been used previously by practitioners for segmentation of soot, e.g., Ref. [11, 14, 26]. Unfortunately, the use of a uniform threshold for the entirety of an input image, often results in misclassifications due to local brightness changes caused by image artifacts, noise, or non-uniform illumination. To solve these problems, adaptive thresholding methods incorporate the local context of a pixel in the decision, whether the pixel belongs to the sought-after object class or not, with a range of variants available [24], though these have seen limited utility in classifying soot TEM images, largely due to the presence of noise in the images.

Many of the current state-of-the-art segmentation methods within the broader field of image analysis are based on machine learning [27]. In addition to using the local context of a pixel, they autonomously learn features that enable them to differentiate objects that belong to different classes, even if they feature similar intensity profiles. This property makes these approaches especially interesting for classifying TEM images, where it is impossible to discriminate intensity variations caused by material contrast, other changes in the electron path, and noise.

In this work, we employ a class of convolutional neural networks (CNNs) as an aggregate-level classifier to segment the aggregates from the background. The procedure was inspired by the semantic segmentation branch of the recently published Panoptic Feature Pyramid Network by Kirillov et al. [28], based on an implementation by Yakubovskiy [29], and realized in PyTorch [30]. We train the network on images of soot collected from a natural gas, internal combustion engine [31]; laboratory gas flare [21]; and marine engine [22, 32]. We then compare the result with several other classifiers available in the literature – including an Otsu classifier [11]; a *k*-means classifier [33]; a trainable Weka classifier [16] – noting the limitations and advantages of each of the techniques. We also compliment aggregate-level classification by considering the sensitivity of a range of fully-automated primary particle sizing techniques to these different classifiers, comparing the primary particle size predicted by PCM [11], EDM-SMS [12], the Hough transform-based method of Kook et al. [15], and the EDM-WS method [13] across a range of classifiers.

## 2. Methods

The goal of aggregate-level segmentation is to produce binary masks where the aggregate or foreground pixels are set to *true* in the corresponding segmentation mask, while the background pixels are set to *false*. Within this work, we chiefly present a convolution neural network (CNN) route to finding this mask, describing the architecture, training, and post-processing used to build the approach.

## 2.1 Aggregate-level classification: Using a CNN

### 2.1.1 Architecture

CNNs designed for semantic segmentation consist of three components: encoder, decoder, and semantic segmentation head. These components are shown schematically in Figure 1.

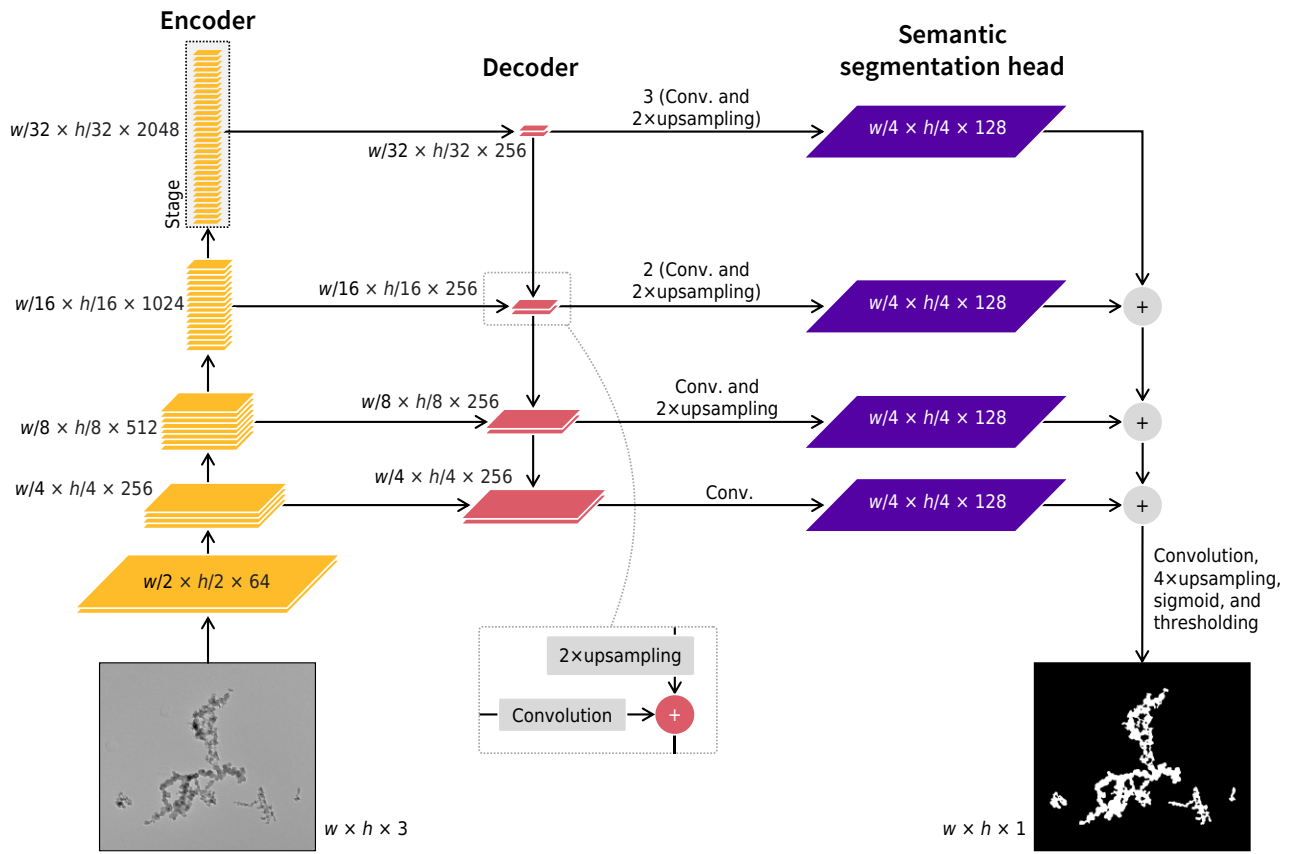

**Figure 1:** Architecture of the convolutional neural network (CNN), with the (left) encoder, which downsamples the image into a compact representation; (middle) decoder, which involves upsampling and convolution with encoder stages; and (right) semantic segmentation head components. Each block in the figure corresponds to a *stage*. In the figure, + denotes *element-wise addition*, which is used to both sum a convolution of an encoder stage with the upsampled output from the previous decoder step as well as to merge the stages of the semantic segmentation head, before transforming them into the final binary mask. *Conv.* is used as a short form for *convolution*. Adapted from Ref. [49].

The *encoder* reduces the spatial resolution of the input via downsampling, while preserving information in a compact representation and discarding noise and other artifacts. Feature extraction is carried out by eponymous convolutional layers, which are composed of a series of *convolution kernels* (a.k.a. *filters*) that calculate an output pixel value as the weighted sum of the values of the corresponding input pixel and its neighborhood pixels. Image convolutions are present in traditional image analysis techniques. A Sobel filter, for example, involves applying two 3x3 convolutional kernels (or perhaps stencils) with fixed weights that approximate gradients in the image (one kernel for the gradient in each of the horizontal and vertical directions), which is useful for edge detection. Unlike traditional filters, however, neural networks automate the choice of convolutional kernel weights, effectively designing optimal filters for a specific task. Encoding often includes maximum pooling layers, which are interlaced with the convolutional layers. During maximum pooling, the input of the respective layer is split into *pools* of a certain size (e.g., 2×2) and only the maximum value of each pool is kept as output. This way, the resolution is halved in both spatial directions and only the strongest features remain. In a feature pyramid network, blocks or *stages* – consisting of multiple convolutional layers followed by a single max pooling layer – are arranged in a way so that the resulting outputs of the blocks have the shape of a pyramid (see Figure 1, left side).

Next, the *decoder* (see Figure 1, middle) increases the spatial resolution of the encoded image by combining (*i*) upsampling, using nearest neighbor interpolation, and (*ii*) convolutions of the encoder stages. To make use of spatial information from different resolutions, the outputs of all stages of the encoding pyramid are used for the decoding. The information is incorporated by adding the convolved output of each encoding block to the input of the corresponding decoding block, i.e., the output of the previous decoding block (see Figure 1, inset).

Finally, the semantic segmentation head (see Figure 1, right side) is used to transform the output of the different decoding stages into the desired output format, in this case a binary mask. As the segmentation head expects a consistent size for processing, the decoder outputs are (*i*) upsampled, again using nearest neighbor interpolation(s) and/or convolution(s); (*ii*) merged by elementwise addition; and, finally (*iii*) convolved and bilinearly upsampled again to match the desired output resolution, i.e., the original resolution of the input image. A sigmoid function is used to map the output for each pixel onto the range (0, 1), which represents the probability that any given pixel belongs to the foreground. Ultimately, pixels featuring probabilities above a threshold of 0.5 are considered foreground pixels.

*2.1.2 Training*

Input data is taken from soot samples collected from a range of sources, including a natural gas, internal combustion engine [31]; marine engine [22]; and a laboratory-scale flare [21], with the number of images from each source indicated in Table 1. The necessary ground truths[1] were produced using the semi-automated *slider* method associated with Dastanpour et al. [11] and updated by Sipkens and Rogak [33]. The method involves (*i*) cropping small regions of the image, (*ii*) adjusting the background for this selected region of the image, (*iii*) applying Gaussian smoothing to the image, and, finally, (*iv*) manually adjusting the threshold in the selected region using a *slider* GUI. This is repeated until each aggregate in the image is characterized and can also be repeated to apply a threshold to small parts of an aggregate that may be missed by larger-scale thresholding. This latter feature allows for user intervention, which, while tedious, generally allows for more accurate classification. Applying a Gaussian smoothing step does tend to produce smoother edges in segmentations but is considered a reasonable injection of prior information given the uncertainties in classifying the boundary pixels in the presence of noise. The binary masks were produced by multiple users in connection with the original studies. The available data was inspected to ensure sufficient quality, ensuring that the ground truth images included aggregates at the borders (while it is a common practice to remove border aggregates, as they would skew subsequent analysis, training requires that all aggergate pixels are labeled) and were not missing parts of the aggregates. The training set did ignore some small, likely spherical aggregates (cf. Figure 4*e* later in this work) in some of the images, in favor of the larger, more complex aggregates in which the original practitioners were more interested. The set of remaining samples was randomly split into training (85%, 345 samples) and validation (15%, 60 samples) image sets.

---

[1] In machine learning, the ground truth – while not necessarily perfect – is the best available data to test predictions of an algorithm.

**Table 1:** Number of images from each soot source used in the training and validation sets, totaling 405 images.

| Soot source | Natural gas, internal combustion engine [31] | Marine engine [22] | Laboratory-scale flare [21] | *Total* |
|---|---|---|---|---|
| Training | 158 | 33 | 154 | *345 (85%)* |
| Validation | 27 | 6 | 27 | *60 (15%)* |
| *Total* | *185 (46%)* | *39 (10%)* | *181 (45%)* | *405* |

Consistently with Kirillov et al. [28], performance while training was assessed using a loss function, $\mathcal{L}$, which compared a third-order tensor composed of a batch of predicted segmentation masks, $\mathbf{Y}^{\text{pred}}$, to the equivalent set of target or ground truth masks, $\mathbf{Y}$, using the pixel-wise binary cross entropy function [34],

$$\mathcal{L}(\mathbf{Y}, \mathbf{Y}^{\text{pred}}) = \frac{1}{Nwh} \sum_{i=1}^{N} \sum_{j=1}^{w} \sum_{k=1}^{h} \left[ Y_{ijk} \log Y_{ijk}^{\text{pred}} + (1 - Y_{ijk}) \log (1 - Y_{ijk}^{\text{pred}}) \right] \quad (1)$$

where $N$ is the number of images in the current training batch; $w$ and $h$ are the image width and height, respectively; $i$ is the image index within the batch; and $j$ and $k$ are the indices of the column and row of the current pixel. Binary cross entropy is designed for the comparison of probability distributions and therefore the de facto standard loss function for classification tasks. This also makes it a reasonable choice for semantic segmentation, which is fundamentally a pixelwise classification. During training, batches of images are segmented by the CNN, and the weights of the CNN layers are adjusted using the Adam optimizer [35].

The learning rate schedule was based on the well-established 3× training schedule [36, 37] but was adapted to the simpler learning task by reducing the step size and training duration. Figure 2 depicts the learning rate schedule. The base learning rate is $\alpha = 10^{-4}$. However, the training begins with a warm-up period of 1,000 iterations, during which the learning rate is increased linearly from $0.001\alpha$ to $\alpha$. After 2,000 and 8,000 iterations respectively, the learning rate is reduced by a factor of 10. The training ends after 10,000 iterations. A complete overview of all training hyperparameters can be found in Table S1. Training was also sped up using transfer learning, initializing the weights of the encoder with weights of a ResNet-50 CNN, which was trained on the ImageNet data set [38, 39]. However, as the images here are very different from the images of the ImageNet data set (i.e., those images were not images of aggregates), the weights of the first encoder layers were not frozen (i.e., all of the node weights were optimized) during the training, such that even the basal weights of the first encoder layers were optimized.

The training was carried out on a dedicated GPU server (see Table S1 and S2). The parameters used for training are summarized in Table 2. The source code and data have been made available online[2].

**Table 2:** Parameters used to train the convolutional neural network.

| Solver | Base learning rate | Warm-up factor | Warm-up period | Learning rate drop steps | Learning rate drop factor | Training duration | Batch size |
|---|---|---|---|---|---|---|---|
| Adam [35] | $10^{-4}$ | 0.001 | 1000 | 2000, 8000 | 0.1 | $10^4$ | 24 (6 per GPU) |

---

[2] The source code of the CNN, the final model and the data sets, used for its training and validation, are available at: https://github.com/maxfrei750/CarbonBlackSegmentation/releases/v1.0. The training and test data sets are also part of the BigParticle.Cloud (https://bigparticle.cloud). The validation image set is included in the Supplemental Information.

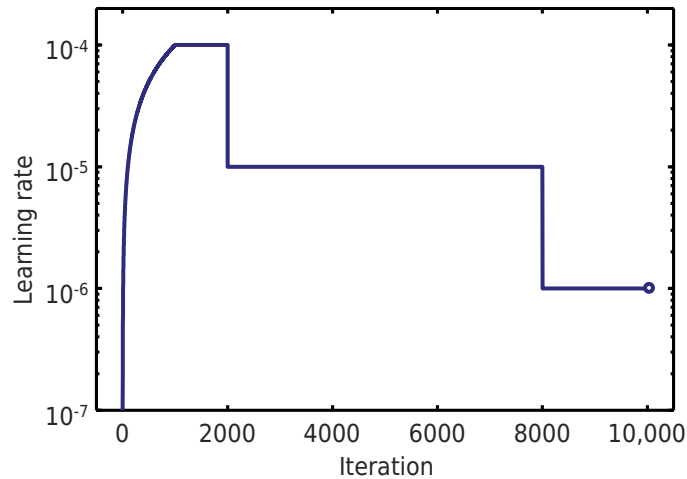

**Figure 2:** Learning rate schedule. Note that the *y*-axis is on a log scale, indicating orders of magnitude changes in the learning rate.

### *2.1.3 Post-Processing*

We also considered additional post-processing using a rolling ball (RB) transform, specifically the variant used for the *k*-means method by Sipkens and Rogak [33] (which itself is a modified version of that proposed by Dastanpour et al. [11]). The size of the element used in the transform was chosen automatically to avoid the requirement of user intervention. We first close the mask with a disc element having a size of 3.2/*pixsize* followed by opening with a disc element having a size of 3.2/*pixsize* – 1, where *pixsize* is the size of a single pixel in the image in nm. This acts to bridge aggregates that may be erroneously fragmented, but where the fragments remain within a certain number of pixels. The effect of this step is considered in Section 3.1.2. We also removed any particles below 1,000 px, which were often erroneous features extracted from the background texture.

### 2.2  Aggregate-level classification: Other methods

We compare these results with segmentations by several other automated methods, including: (*i*) Otsu thresholding combined with background subtraction and a RB transform, as first proposed by Dastanpour et al. [11] and later improved by Sipkens and Rogak [33]; (*ii*) trainable Weka segmentation (TWS) via Fiji [40], following from Altenhoff et al. [16]; and (*iii*) the *k*-means approach of Sipkens and Rogak [33]. For TWS, this training presents some differences relative to the original study of Altenhoff et al. [16] and the study of Sipkens and Rogak [33], including that a single classifier was trained for all the images and that the training set data used a higher fraction of pre-classified pixels, which increased the detail in the labels relative to the previous reapplication. Overall performance remained similar to previous studies.

## 2.3 Primary particle sizing

Primary particle sizing is evaluated from the segmented aggregates using four, previously-established techniques. In all cases, there exist small changes relative to the original works, with the methods unified into a single Matlab software package[3] to facilitate rapid, open comparison.

We employ the pair correlation method (PCM), adapted from Dastanpour et al. [11]. Changes were very minor relative to the previous implementation, mostly associated with improvements in memory usage within the original algorithm and a decoupling of the aggregate-level classifier provided in that software distribution.

We also adapted the Euclidean distance mapping, surface-based scale analysis method (EDM-SBS) from Bescond et al. [12] into a Matlab equivalent, decoupling the initial ImageJ classifier associated with that work. While this may cause some backwards compatibility issues, the overall nature of the method remains true to the original work. First, a Euclidean distance transform is applied to the binary mask. The transformed image is then eroded one level at a time, counting the number of remaining pixels at each level. This curve is transformed to a common discretization in nanometers and then normalized before being fit with a sigmoid function of the form

$$\frac{S(d)}{S(0)} = a \left\{ 1 + \exp\left[ \frac{\left(\ln d - \ln d_{p,g}\right)/\sigma_{p,g} - \beta}{\Omega} \right] \right\}^{-1}, \qquad (2)$$

with calibration constants $a = 0.9966$, $\beta = 1.9658$, and $\Omega = -0.8515$ from the code [41] associated with the aforementioned paper (noting a minor discrepancy between the values used in the code and those proposed in the published work, $a = 1.0$, $\beta = 1.9$, and $\Omega = -0.8$, and that the level of precision in the former is over specified relative to the expected uncertainties). This procedure was applied both on the full set of binary masks simultaneously and at an aggregate-level, which differs from the previous implementation. The smaller number of pixels in each aggregate may limit the accuracy of the per-aggregate statistics (that is, statistics computed for each aggregate, rather than for individual primary particles inside the aggregates or for the ensemble of aggregates more generally) but allows comparison against the other methods, which do output this kind of information.

We also adapt the Hough transform method from Kook et al. [15]. We apply the same processing steps and Matlab's *imfindcircles* function, looking for circles in the range of 16 to 100 px in diameter. We then assign the circles to their respective aggregates using the binary masks, which also acts to automatically remove any erroneous circles found in the background. Unlike the previous two methods, this approach can estimate the location of specific primary particles, rather than simply stating an average primary particle diameter for each aggregate.

Finally, we adapt the Euclidean distance mapping-watershed method (EDM-WS) of De Temmerman et al. [13]. As with EDM-SBS, we start by applying the Euclidean distance transform to the binary mask. This time, however, we use the watershed transform on the EDM result, after first removing local minima with a prominence of less than a single level. Theoretically, and similar to the method from Kook et al. [15], this method can spatially locate specific perceived primary particles.

---

[3] The *atems* package available at https://github.com/tsipkens/atems. This package includes implementations of the Otsu and *k*-means classifiers discussed elsewhere in this work and features other analysis tools used to interpret the results (e.g., the primary particle sizing techniques and aggregate matching tool).

# 3. Results

## 3.1 Aggregate-level segmentation with the CNN

### 3.1.1 Validating the raw CNN output

Qualitatively, the CNN robustly identifies aggregates in most images. Sample segmentations are shown in Figure 4, alongside the validation data from the slider method and other automated methods discussed in Section 3.2. We quantified the accuracy of the CNN itself, without considering the physical consequences, using two metrics.

Perhaps the simplest is the confusion matrix, using the raw output from the CNN for the validation images to categorize pixels into true aggregate pixels, $T_{\text{agg}}$; false aggregate pixels, $F_{\text{agg}}$; true background pixels, $T_{\text{bg}}$; and false background pixels, $F_{\text{bg}}$. By this measure, the CNN does well (cf. Table 3 for comparison to other methods), misidentifying only 0.09% of the pixels from the background and 3.2% of the pixels from the aggregates relative to the ground truth provided to the network. An overall measure of the accuracy is computed from these quantities as [42],

$$\text{Accuracy} = \frac{T_{\text{agg}} + T_{\text{bg}}}{T_{\text{agg}} + T_{\text{bg}} + F_{\text{agg}} + F_{\text{bg}}}. \tag{3}$$

corresponding to 99.8% for this case.

For semantic segmentation, a second common measure is the intersection-over-union (IoU). This quantity is computed as the ratio of pixels that are classified as aggregate by both the validation images and the CNN (the intersection of the sets) to those that either model described as aggregate (the union of the sets). The IoU has an optimal value of IoU = 100%, where the masks are identical. The CNN again performed well here, with an IoU = 94%.

### 3.1.2 Assessing post-processing steps for the CNN

While the above metrics are useful in terms of evaluating the basic accuracy of the CNN, these metrics fail to capture other characteristics of the segmentation that are paramount to fully automating aggregate characterization. One of the largest problems occurs when aggregates are fragmented, such that a fully automated method would distribute a single aggregate's characteristics over multiple *perceived aggregates*. In this case, while the number of misclassified pixels may be quite small, the consequences for automated aggregate characterization are significant. Observations suggest that this could occur as a fragmenting of the peripheral components of the aggregate (e.g., Figure 3a) or as the breaking up a larger aggregate when two portions of the aggregate are only connected by a narrow region of pixels. Inspection shows that this does occur in the raw CNN segmentations, partially evidenced by a significant increase in the number of perceived aggregates, from 164 for the slider method to 199 identified by the CNN (cf. Table 3 in connection with the comparison of classifiers in Section 3.2).

To address this issue, we apply a rolling ball transform, as described in Section 2.1.3. The overall effect of the transform on a sample CNN classification is demonstrated in Figure 3, combining four fragments of the larger aggregate properly into a single object. It is worth noting that the way the rolling ball transform connects fragments is not *smart*, i.e., the method has no knowledge of the underlying image and whether darker, aggregate-containing pixels actually connect two fragments. However, the advantages of connecting these fragments outweighs any minor artifacts that may be introduced by this process. In fact, the IoU decreases very little (percent difference of -0.17%) when adding the transform to the CNN results.

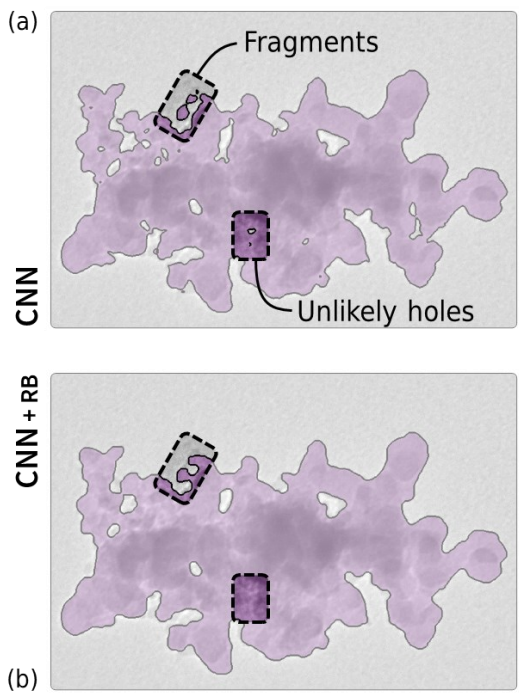

**Figure 3**: The effect of using the rolling ball (RB) transform demonstrated for a single aggregate. The output of the CNN, (a), has several unlikely holes and stray fragments that have been resolved after applying the rolling ball transform, (b).

After the rolling ball transform, several small, perceived aggregates remain incorrectly identified, often erroneously capturing texture in the background. This was resolved by removing any perceived aggregates less than 1,000 px in size, chosen by inspection. While this does present a lower limit on the size of aggregates that can be classified by the method, it reduces the number of perceived aggregates to within 10 of the slider classifier used as the ground truth. This step also increases the IoU, albeit very slightly (percent difference of +0.04%). Subsequent discussions of the "CNN + RB + 1,000 px" variant of this classifier combine the rolling ball step with this small aggregate removal procedure.

The overall effect of introducing the rolling ball and small aggregate removal steps is compared for a larger range of metrics (with a comparison to other methods, cf. Section 3.2) in Table 3. Beyond the overall reduction in the number of particles, attempts to match the aggregates between the CNN and slider classifiers[4] shows that the reduction in the number of identified aggregates is also associated with a significant improvement in the aggregates that can be matched between the slider and CNN classifiers.

---

[4] As mentioned previously, code to match aggregates between to aggregate sets is included in the *atems* package, specifically the function available at: https://github.com/tsipkens/atems/blob/master/%2Btools/match_aggs.m. Matches are roughly assessed by considering the center of mass of the aggregates (considering centers within 200 px) and, for larger aggregates, overlap (requiring more than 30% pixel overlap for aggregates larger than 30,000 px).

**Table 3:** Parameters assessing the various aggregate-level classifiers. False positives, false negatives, accuracy, and intersection-over-union (IoU) are taken with respect to the slider segmentation, which is considered the *ground truth*, and neglect images where a given method was automatically detected to have failed. Bracketed values in the IoU column correspond to the more direct IoU, where images on which a classifier fails are included in the statistics and treated as if they do not contain any aggregates. The IoU and accuracy consider the segmentations prior to analysis of the binary mask to identify and remove border aggregates. Accuracy, false positives, and false negatives are only reported for the original classifiers, prior to additional processing. Added and removed aggregates are determined by attempting to match aggregate centers and pixel overlap. The sum of the aggregates added and removed is the difference between the number of aggregates for a given method and the number of aggregates for the slider method. The percent difference in the projected area-equivalent diameters, $\Delta d_{a,m}$ and $\Delta d_{a,g}$, corresponds to the difference in the geometric mean and median, respectively, computed across the particles identified by each method, thus yielding estimates of accuracy in terms of computing the central tendency of the aggregate size distribution.

| Method | False positive | False negative | Accuracy | No. of failed images | IoU | No. of aggs. | Added aggs. | Removed aggs. | $\Delta d_{a,m}$ | $\Delta d_{a,g}$ |
|---|---|---|---|---|---|---|---|---|---|---|
| Slider (*ground truth*) | - | - | - | - | - | 158 | - | - | - | - |
| **CNN** | | | | | | | | | | |
| Original | 0.09% | 3.2% | 99.8% | 0 | 94% | 199 | +41 | 0 | -14% | -40% |
| + RB | - | - | - | - | 94% | 182 | +24 | 0 | -9.0% | -22% |
| + 1,000 px | - | - | - | - | 94% | 168 | +10 | 0 | -1.4% | -6.6% |
| + RB + 1,000 px | - | - | - | - | 94% | 166 | +9 | -1 | -0.43% | -4.5% |
| ***k*-means** | | | | | | | | | | |
| Original (incl. RB) | 0.22% | 6.6% | 99.5% | 11 | 89% (23%) | 211 | +75 | -22 | -20% | -45% |
| + 1,000 px | - | - | - | - | 89% (23%) | 165 | +29 | -22 | -0.01% | -9.2% |
| Otsu + RB | 0.05% | 21% | 99.0% | 18 | 78% (11%) | 260 | +139 | -37 | -39% | -48% |
| TWS (WEKA) | 0.33% | 5.8% | 99.5% | 2 | 87% (86%) | 205 | +59 | -12 | -10% | -22% |

## 3.2 Cross-comparison of multiple classifiers

A range of automated, classifiers were run on the 60 members of the validation image set. Figure 4 shows the resultant binary masks overlaid on the original images for a subset of the validation images, after removing boundary aggregates and adding a circle indicating the projected area-equivalent diameter about the aggregate center-of-mass. Figure 5 shows corresponding kernel density estimates of the projected area-equivalent diameter distribution for the range of classifiers. Overall, all of the considered methods, except the Otsu approach, performed reasonably, consistently capturing the primary mode of the projected area-equivalent diameter distribution.

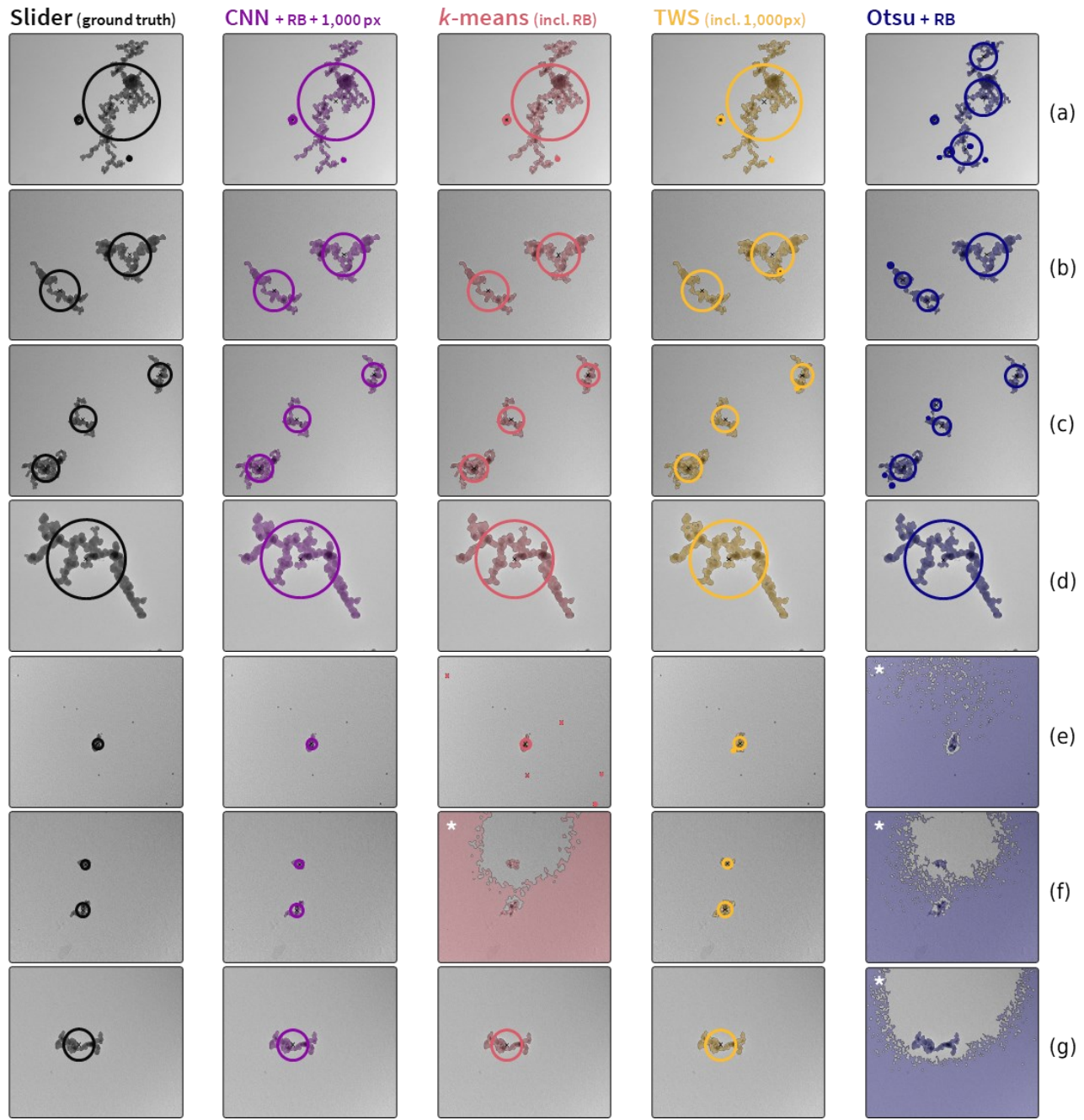

**Figure 4:** Segmentations for a random subset of the validation data set, using, from left to right: (i) the slider method, i.e., the *ground truth*; (ii) the CNN with a rolling ball transform; (iii) the *k*-means approach of Sipkens and Rogak [33], with the default aggregate size threshold and which included the rolling ball transform described in that work; (iv) Otsu thresholding with a rolling ball transform, following Dastanpour et al. [11]; and (v) trainable Weka segmentation, via Fiji [40] and following Altenhoff et al. [16], though using a different training. Items marked with "*" represent catastrophic failures of a method, and the raw segmentations are shown in place of characterized images (e.g., panel (*f*) for *k*-means). The TWS (WEKA) method also occasionally failed, resulting in a speckling of small, perceived aggregates in the background, though the failure was not noted in the random subset shown here (the failure didn't overlap with the *k*-means and Otsu failures). Panel (*e*) features several small, likely spherical particles that were identified by the k-means method but were ignored in the *ground truth* and other methods.

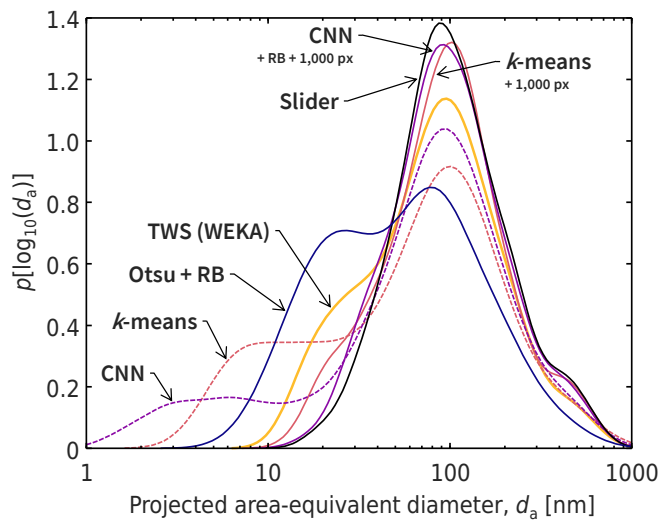

**Figure 5**: Kernel density estimates of the projected area-equivalent diameter distribution for a range of aggregate-level classifiers. Distributions are specified in $\log_{10}(d_a)$ space, such that a lognormal distribution would appear Gaussian. Dashed lines correspond to the original CNN and *k*-means classifiers, without adding the 1,000 px aggregate size threshold or the rolling ball (RB) transform for the CNN.

We note that the Otsu and *k*-means classifiers occasionally fail rather catastrophically, consistent with observations by Sipkens and Rogak [33]. In Figure 4, aggregate characterizations for these cases are replaced with the underlying classifications and marked with an asterisk. While the code associated with the *k*-means and Otsu methods[5] can automatically identify and neglect these images in computing aggregate statistics, the CNN is advantageous in that these kinds of failures did not occur in any of the validation images. Otsu failed about 70% more often than *k*-means. The observation that one of the problem images occurred in both the *k*-means and Otsu classifiers likely stems from the *k*-means classifier using Otsu as a starting point for the *adjusted threshold layer* in the overall *k*-means procedure.

While not shown in the randomly sampled data set, the TWS results also failed on occasion (two images in the validation set), in this case featuring results with a speckling of small perceived aggregates across the entire image, including the background. As with Otsu and *k*-means, the post-processing code can find and neglect these images.

Table 3 shows the number of perceived aggregates predicted by the various methods, alongside other metrics. Initial *k*-means estimates predicted a significantly higher number of aggregates and much smaller geometric mean projected area-equivalent diameter than the slider method. From inspection, we noted that many of these additional perceived aggregates correspond to real, physical features in the images, specifically small, presumably spherical particles collected from the flare [21] (several are identified in Figure 4*e*). Other diagnostics would likely be required to identify the composition of these particles, including whether they are truly soot or some other kinds of particles (e.g., microscopist have noted salt, lubricating oil, or other contaminants mixed with soot on TEM grids [32, 43]). The relative optical depth method of Tian et al. [17] may also be able to provide some information here but requires some indirect inference and currently requires user intervention to calibrate the method. Overall, the original *k*-means classifier [33] seems capable of resolving aggregates down to ~50 px. These particles were ignored when the slider method was first applied by the authors of the original studies. Naturally, since the training data ignored these particles, the CNN is incapable of identifying these particles. These particles currently lie below the 1,000 px threshold applied to remove artifacts during CNN post-processing, a limit set to avoid the CNN from perceiving textured parts of the background as aggregate. More work would be required to examine if the CNN could be trained to identify these smaller particles, adjusting the training data and weights to favor thin aggregate sections. Even then, it seems probable that the *k*-means classifier will have a lower detectability limit. If one adds the same 1,000 px threshold to the *k*-means procedure (marked with "+ 1,000 px" in Table 3 and Figure 5), there is a significant improvement in consistency between the number of identified aggregates and the geometric mean projected-area equivalent diameter relative to the slider method. For these larger particles, the CNN generally outperforms the *k*-means, not falling victim to the same failures noted in connection with Figure 4, exhibiting a better IoU, and exhibiting more consistency with the slider results in terms of the number aggregates (including reductions in the number of aggregates without a suitable match).

---

[5] Functions for the Otsu and *k*-means methods are available in the *atems* codebase hosted at https://github.com/tsipkens/atems.

These observations have consequences for the projected area-equivalent diameter distributions in Figure 5. Both the original binary output from the CNN and the original *k*-means procedures feature a heavy, lower tail. However, the preceding discussion provides very different causes for this tail between the two classifiers. The original output from the CNN features many false aggregates in the textured background, which will occur in this tail. As a result, the tail continues to smaller particle sizes and features a very minor, second peak centered around 3 nm, which is around the average magnitude for features in the textured background of the images. The *k*-means classifier, in contrast, captures real features in the image at these smaller sizes. In both cases, the primary mode for the larger aggregate sizes is quite consistent, both in terms of their central tendency and distribution width. Unsurprisingly, then, when these small particles are removed by ignoring perceived aggregates below 1,000 px (i.e., using the "CNN + RB + 1,000 px" and "*k*-means + 1,000 px" classifiers), the projected area-equivalent distributions are very similar, both between the two methods and with the slider result.

Consistent with Sipkens and Rogak [33], the Otsu classifier performed poorly even when it did not fail, often breaking aggregates into multiple fragments, resulting in a significant increase in the number of aggregates and a significant reduction in the median projected area-equivalent diameter. This also results in an artificial, heavy lower tail in the projected area-equivalent diameter distributions. Since the method typically fragmented larger aggregates (e.g., splitting aggregates into two large pieces or fragmenting edge components) instead of adding features from the background, the projected area-equivalent diameter distribution for the Otsu classifier is considerably less skewed than the raw CNN results and features a primary peak at smaller particle sizes than the slider results.

Trainable Weka segmentation (TWS) performed well for most of the images, including those in Figure 4. Relative to Sipkens and Rogak [33] and Altenhoff et al. [16], the retrained TWS resulted in some improvement in the aggregate areas, with less padding around the aggregates that the figures in those works. There is slightly more fragmenting for this image set (e.g., Figure 4*a*) than the simpler image set considered in Ref. [33]. We note that this fragmenting occurred for both the TWS trained in Sipkens and Rogak [33] and that trained for this work, suggesting differences are associated with the larger image set used in this work rather than a consequence of structural difference between the two classifiers. Overall, the TWS performed reasonably in predicting the primary mode of the projected area-equivalent diameter, with a central tendency consistent with the slider, *k*-means, and CNN classifications. These distributions again feature a heavier lower tail, associated with some fragmentation of the aggregate masks and that is considerably less significant than the Otsu result.

The authors also considered adaptive thresholding via Matlab, wherein the threshold was automatically adjusted for different regions of the image. However, poor results (worse than Otsu) were encountered for a wide range of neighbourhoods and sensitivities, such that the method was not considered further in this work.

Computational effort was similar across most of the automated methods. Otsu was substantially faster (reducing computational effort by approximately a factor of 4), though this is likely irrelevant given (*i*) the larger errors (the degree of manual post-processing would more than make up for the faster runtimes) and (*ii*) offline nature of the problem. All of the automated methods were associated with a substantial decrease in the overall processing time relative to the slider method (more than an order of magnitude, with real improvements depending on any number of factors, including user familiarity with the more manual methods) as well as being passive (not requiring active involvement by the user during classification).

### 3.3 Primary particle sizing

With binary masks available, several automated methods can now be used to provide primary particle size information and to assign this information on a per-aggregate basis. For PCM [11], EDM-SBS [12], and EDM-WS [13], the binary mask produced by the aforementioned classifiers are used directly as input. For the Hough transform method, following Kook et al. [15], the original image is used as input to the algorithm and the binary masks are only used to select which of the circles found by the algorithm belong to a specific aggregate (including rejecting circles found in the background). Accordingly, differences between the classifiers are expected to be smaller for the Hough transform method than the other primary particle sizing techniques, due to the nature of the technique, rather than a higher fidelity to the true primary particle diameter. Table 4 shows the ensemble geometric mean and standard deviation of the geometric mean primary particle size per-aggregate for a range of aggregate-level classifier-primary particle sizing technique pairings. Note that the reported ensemble statistics are not the same as those for the underlying primary particle size distribution more generally, as individual primary particle information is not available for PCM and EDM-SBS on a per-aggregate basis. The PCM method also returns a mean primary particle diameter per-aggregate instead of a geometric mean, such that the reported quantity differs from the other methods presented here. This will incur larger discrepancies for cases where individual aggregates contain large changes in the primary particle size, at which point the assumptions underlying PCM begin to breakdown.

We proceed with two considerations for the different pairings of aggregate-level classifier and primary particle sizing method: (*i*) the sensitivity of a given primary particle sizing method to the aggregate-level classifier, addressed by conditioning on the primary particle sizing technique in Figure 6, and (*ii*) structural variations between the primary particle sizing techniques for a given classifier, addressed by conditioning on the aggregate-level classifier in Figure 7.

The former consideration refers to how consistent a given primary particle sizing technique will be if different classifiers are used. To be clear, this will only give an indication of *sensitivity*, rather than a measure of *accuracy*. In other words, this treatment filters away structural errors associated with a given primary particle sizing technique (e.g., this treatment cannot be used to show if the Hough transform gives primary particle sizes that are larger on average) in favor of variations within a primary particle sizing method. Within each group – corresponding to a single primary particle sizing technique – the primary particle size is normalized by that inferred for the slider mask, which is the closest we have to a ground truth. As such, parity in the figure corresponds to primary particle sizes that match the slider results for a given primary particle sizing technique. The authors note that while the slider method was taken as the ground truth within each group, Gaussian smoothing of the image during that procedure will tend to smooth out image edges, such that the aggregate outline remains imperfect. Within each pairing, overlaid data is plotted such that the progression from left-to-right indicates an increase in primary particle size derived using the slider classifier, giving a sense of the structure, if any, in the discrepancy.

Figure 6*a* shows that the primary particle size predicted by the Hough transform method is nearly identical for all of the classifiers. This is unsurprising given the algorithm uses the original image instead of the binary mask, only using the mask to determine which circles are considered as *part* of the aggregate. Some spread is

**Table 4:** Statistics summarizing the primary particle size distribution computed for different binary masks and primary particle sizing techniques. Results are reported as ensemble geometric means (unbracketed) and geometric standard deviations (bracketed) of the geometric mean primary particle size for each aggregate. An exception occurs for PCM, where the method returns the mean primary particle size per aggregate instead of the geometric mean. Even then, PCM ensemble statistics are taken as the "geometric" quantities. Note that the exceptionally high distribution width for the EMD-SBS and CNN (original) pairing is correct and associated with a large number of very small particles with perceived primary particle diameters less than 1 nm. We note that EDM-SBS is expected to be inaccurate for these scenarios, as the sigmoid fitting function is applied far from its original validation point. The Hough method is that corresponding to Kook et al. [15].

| Agg. classifier | Hough [nm] | EDM-SBS [nm] | PCM [nm]* | EDM-WS [nm] |
|---|---|---|---|---|
| Slider (*ground truth*) | 17.4 (1.44) | 19.0 (1.52) | 18.4 (1.52) | 29.2 (1.44) |
| **CNN** | | | | |
|   Original | 17.4 (1.47) | 11.0 (5.41) | 13.8 (2.21) | 20.7 (2.27) |
|   + RB + 1,000 px | 17.6 (1.46) | 19.8 (1.49) | 20.8 (1.45) | 31.2 (1.39) |
| ***k*-means** | | | | |
|   Original (incl. RB) | 16.7 (1.47) | 13.8 (2.42) | 16.3 (1.99) | 24.7 (1.99) |
|   + 1,000 px | 19.9 (0.42) | 20.9 (0.25) | 17.5 (0.14) | 31.3 (0.24) |
| Otsu + RB | 17.4 (1.42) | 12.2 (2.02) | 9.1 (1.87) | 19.6 (1.70) |
| TWS (WEKA) | 17.5 (1.45) | 14.1 (1.83) | 13.4 (1.83) | 20.4 (1.70) |

\* PCM results are the ensemble geometric mean and standard deviation of the mean primary particle diameter per aggregate, whereas other results are calculated based on the geometric mean primary particle diameter per aggregate. However, the size distribution is expected to be rather narrows, such that these discrepancies are likely small.

observed for the Otsu method, which incurred significant fragmenting, changing which circles are included in a given aggregate.

    The other primary particle sizing methods, Figure 6*b-d*, exhibited a more significant dependence on the classifier, as expected. Generally, PCM pairings were the most sensitive to the aggregate-level classifier (easily up or down by a factor of two), while EDM-SBS pairings were the least sensitive (generally resulting in primary particle sizes within 50% regardless of the aggregate-level classifier). Across PCM, EDM-SBS and EDM-WS, Otsu showed significantly larger variations relative to the slider method. Otsu results were also consistently biased downward, corresponding to smaller primary particle sizes on average, likely a consequence of the aforementioned fragmentation of the aggregates. The CNN pairings, by contrast, were the most consistent with the slider method, which is perhaps unsurprising given that the CNN was trained on the slider segmentations.

    By contrast, biases between the different primary particle sizing techniques were assessed using a second set of box-whisker plots in Figure 7. Here, for each aggregate-level classifier, the EDM-SBS method was chosen to normalize the points on the *y*-axis, as it performed consistently in connection with Figure 6.

    In general, the EDM-WS data showed a significant bias upwards (between 50 and 100% on average), consistent with previous observations that the method tended to group together central clumps into single primary particles. Of the different aggregate-level classifiers, the Otsu segmentations resulted in the largest variability in the primary particle size, particularly when paired with the Hough transform, while the CNN and slider methods showed the largest consistencies with EDM-SBS. There was also a large degree of consistency between the slider and CNN results in considering individual particle characteristics, with both methods exhibiting very similar data cloud shape and tail behaviour across all of the primary particle sizing techniques. Similar to Figure 6, we note that, within each pairing, overlaid data is plotted such that the progression from left-to-right indicates an increase in EDM-SBS primary particle size. In this regard, there is also often a clear structure in the discrepancy between EDM-SBS and the other primary particle sizes, with the smaller primary

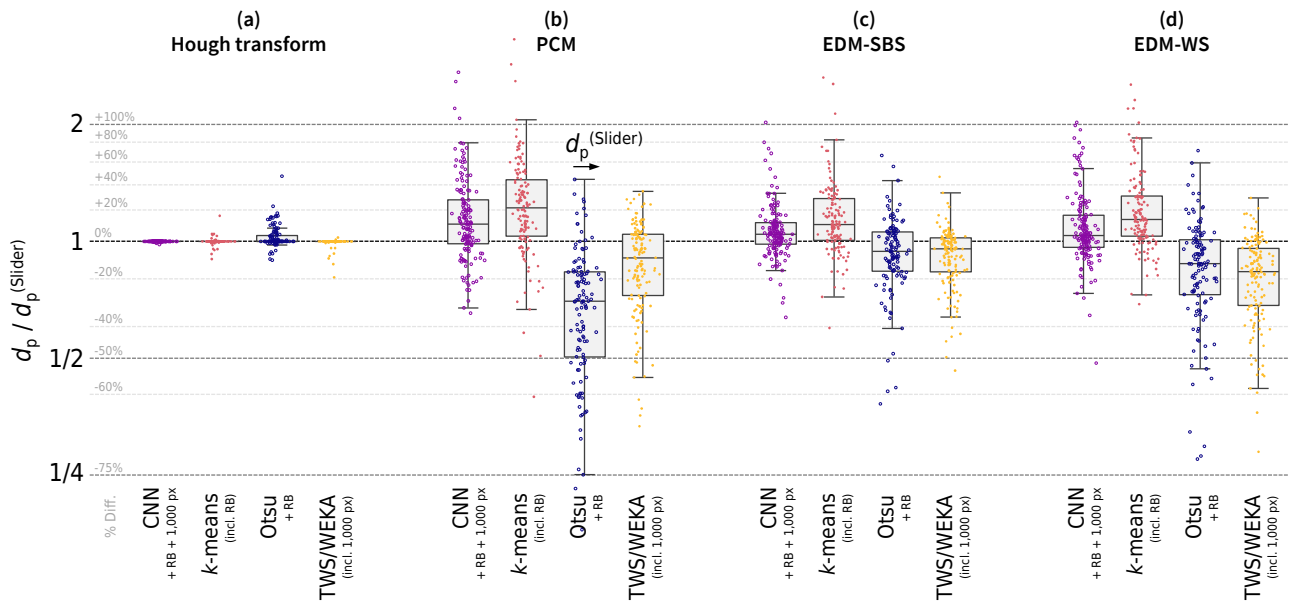

**Figure 6:** Box and whisker plot of the average primary particle size per aggregate, normalized by the primary particle size computed using the slider segmentation for each particle sizing method (i.e., depicting the variation across multiple aggregate-level classifiers within a primary particle sizing method). Within each data cluster, individual points, proceeding from left to right, correspond to increasing primary particle size for the slider classifier on a log-scale. Edges of the box correspond to the 25th and 75th percentiles, and whiskers to 1.5 times the interquartile range. The vertical axis is on a log-scale, such that even increments correspond to an increase or decrease in a multiplicative factor (e.g., 2×$d_p^{(Slider)}$). Smaller increments of percent difference are shown in light gray, so as to not overly busy the figure. Hough transform results, following Kook et al. [15], show minimal variation only because the primary particle size is not sensitive to the aggregate-level classifier, yielding nearly identical results for the different classifiers.

particles more likely to appear higher on the plot (i.e., the other methods typically predicated larger primary particle size when the EDM-SBS primary particle size was small). Overall, the Hough transform primary particle sizes were the least consistent with the EDM-SBS results, often resulting in differences exceeding a factor of two (i.e., percent differences of +100%/−50%).

Across all of the techniques, a subset of which are shown in Figure 8, the primary particle size is observed to increase with projected area-equivalent diameter, consistent with previous observations (e.g., Ref. [44]). This does not add new evidence of this trend, as it was noted in the original studies for much of this data. It does, however, indicate that this trend is robust to a range of analysis techniques. Figure 8 also indicates data ellipses, expected to encompass 67% of the data and that show the degree of correlation between $d_p$ and $d_a$, and implied power law fits corresponding to,

$$d_p = d_{p,100} \left( d_a / 100 \text{ nm} \right)^{D_{TEM}} \tag{4}$$

where $D_{TEM}$ is the slope of the straight lines in Figure 8 and can be related to the mass-mobility exponent, $d_a$ is given in nm, and $d_{p,100}$ is the expected primary particle diameter of an aggregate with $d_a$ = 100 nm. Power law parameters are reported in Table 5. Figure 8 also indicates the universal relation proposed by Olfert and Rogak [19] (solid gray line), where $d_{p,100}$ = 17.8 nm and $D_{TEM}$ = 0.35, derived for a range of sources, largely using the slider-PCM pairing.

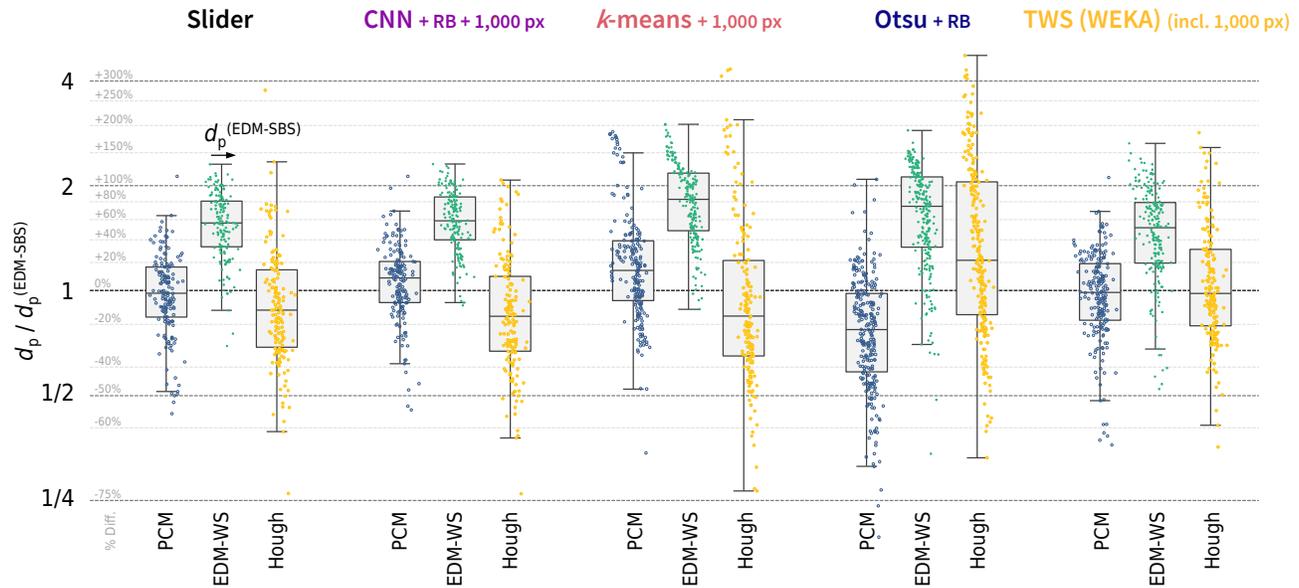

**Figure 7:** Box and whisker plots of the average primary particle size per aggregate for each aggregate-level classifier, with primary particle sizes phrased relative to the EDM-SBS method for each classifier. Within each data cluster, individual points, proceeding from left to right, correspond to increasing EMD-SBS primary particle size on a log-scale. Edges of the box correspond to the 25th and 75th percentiles, and whiskers to 1.5 times the interquartile range. The vertical axis is on a log-scale, such that even increments correspond to an increase or decrease in a multiplicative factor (e.g., $2 \times d_\text{p}^{(\text{EDM-SBS})}$). Smaller increments of percent difference are show in light gray.

Data ellipses take on a similar shape for all of the pairings. Once again, the EDM-WS method is shown to predict larger primary particle sizes, with Figure 8 explicitly showing that this occurs for the same projected area-equivalent diameter (a consequence of the same binary mask being used in each panel). The power law fits are generally distributed about the universal relation of Olfert and Rogak [19]. The slope varies between the different classifier-primary particle sizing technique pairs, likely indicative of larger uncertainties in $D_\text{TEM}$. Overall, $D_\text{TEM}$ is distributed about 0.32, with a standard deviation across the various pairings (an estimate of the uncertainties associated with the image analysis routine) of 0.14. However, the differences are highly structured with the primary particle sizing technique. The Hough transform pairings universally predict considerably smaller $D_\text{TEM}$, while the EDM-SBS pairings always incur a larger $D_\text{TEM}$. The PCM and EDM-WS incur nearly the identical slopes for any given classifier but have slopes that vary between classifiers. Overall, naïve uncertainties (taken as if the user had no knowledge of which method was preferred a priori) in $D_\text{TEM}$ due to the chosen classifier-primary particle sizing pairing are estimated at 90%. All of this casts a caution on the reliability of slope derived from this data. Improving precision would likely require a larger cross-laboratory study of the range of classifier-primary particle sizing technique pairings, including manual and semi-automatic (e.g., CRES [18]) primary particle sizing, to correctly identify which automated methods are more reliable in predicting this quantity. What we can conclude, however, is that there is a consistent positive correlation, regardless of the image analysis technique used, even for the EDM-WS pairings with its different underlying representation of primary particle size. This adds significant support for the theory that larger aggregates typically have larger primary particles, with consequences for their optical and transport properties.

In contrast to $D_\text{TEM}$, the value of $d_\text{p,100}$ was reasonably consistent between all but the EDM-WS pairings. The results here suggest a cross-method value of $d_\text{p,100}$ = 18.5 nm, with a standard deviation across the various pairings (excluding EDM-WS) of 1.9 nm (resulting in ~10% error, considering two standard deviations). This is consistent with the universal relation of Olfert and Rogak [19], who suggested 17.8 nm.

It is worth noting that most of the methods considered in this work use some form of rolling ball transform or morphological operation to filter the final image, placing them within a specific class of methods. This feature may result in an underestimate of the uncertainties associated with the image analysis approach when considering methods that deviate from this class.

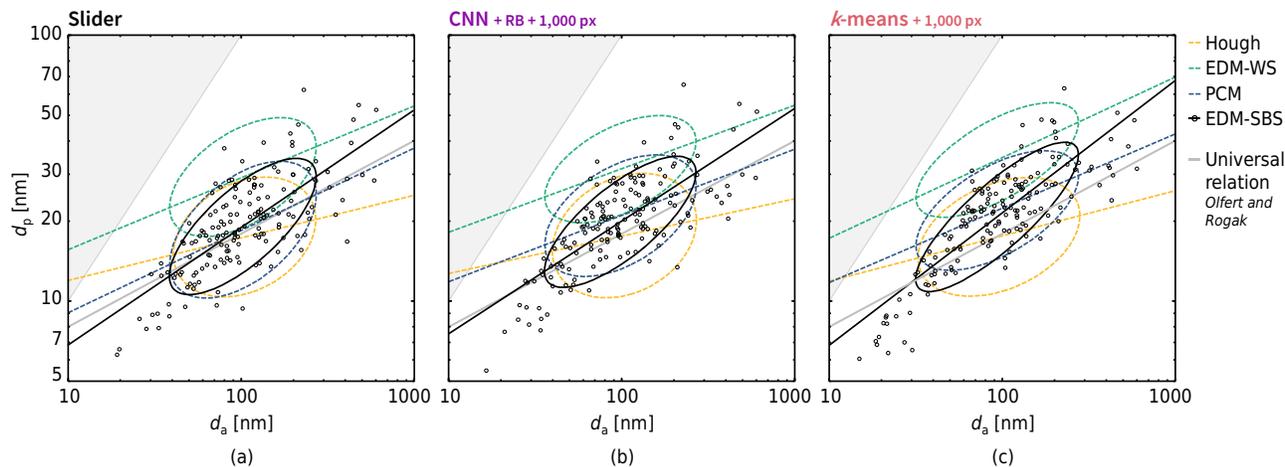

**Figure 8:** Trends in the average primary particle per aggregate with projected area-equivalent diameter, with 1σ data ellipses and power law fits. The data cloud is shown for the EDM-SBS cases for demonstration. The solid, gray line corresponds to the universal relation from Olfert and Rogak [19]. The shaded corner in the upper left corresponds to scenarios where the primary particle would exceed the projected area-equivalent diameter, which are unphysical.

**Table 5:** Power law fits for the relationship between the projected area-equivalent diameter and average primary particle diameter for different classifier-primary particle sizing technique pairings. For each pairing, parameters in not in brackets correspond to $d_{p,100}$, while parameters in brackets are $D_{TEM}$. Average and standard deviation are taken over the classifiers for the same primary particle sizing technique. Otsu was excluded here, as it was shown to underperform previously in this work.

| Agg. classifier | Hough (Kook et al.) [nm] | EDM-SBS [nm] | PCM [nm] | EDM-WS [nm] |
| --- | --- | --- | --- | --- |
| Slider (*ground truth*) | 17.2 (0.16) | 18.8 (0.44) | 18.3 (0.31) | 28.9 (0.27) |
| CNN + RB + 1,000 px | 17.5 (0.14) | 19.9 (0.42) | 20.9 (0.25) | 31.3 (0.24) |
| k-means + 1,000 px | 17.5 (0.17) | 21.4 (0.50) | 22.2 (0.28) | 34.5 (0.30) |
| TWS (WEKA) | 17.5 (0.12) | 16.1 (0.60) | 14.9 (0.47) | 22.4 (0.42) |
| Average | 17.4 (0.15) | 19.1 (0.49) | 19.1 (0.33) | 29.3 (0.31) |
| Standard deviation | 0.2 (0.02) | 2.2 (0.08) | 3.2 (0.09) | 5.1 (0.08) |

## 4. Conclusions

The current work has presented steps towards fully automating the process of characterizing soot in TEM images. Results were reasonable for the CNN, *k*-means, and TWS classifiers, producing very similar segmentations and projected area-equivalent diameter distributions. The *k*-means classifier was observed to have a lower detectability limit, down to 50 px, but failed entirely on some images (a little above 15% for the considered image set). In contrast, while the CNN underperformed for the few particles below 1,000 px (such that these aggregates were removed in a post-processing step), the classifier outperformed all of the other classifiers for a vast majority of the aggregates. The results showed a high degree of fidelity to the results of the semi-automated slider analysis used as the ground truth, with an accuracy of 99.8%; an IoU of 94%; an error in the median and geometric mean projected area-equivalent diameters of less than 2% and 5%, respectively, for this data set; and a matched aggregate count within 2%. Overall, the automated methods are expected to result in more than an order-of-magnitude improvement in terms of classification time.

Primary particle sizing techniques were generally found to be sensitive to the chosen classifier, resulting in uncertainties in the primary particle size that significantly exceed those in the projected area-equivalent diameter. Of the methods that take a binary mask as an input, EDM-SBS was shown to be the least sensitive to the classifier (which is not to say it is necessarily the most accurate). Without any knowledge of the true primary particle size, uncertainties are estimated at around a factor of two (i.e., +100%/−50%).

The EDM-SBS, PCM, and Hough transform (following Kook et al. [15]) all show some form of positive correlation between primary particle size and projected area-equivalent diameter and predict very similar values for $d_{p,100}$, around 18.4 nm. This adds significant support to the theory that the primary particle size scales with overall aggerate size, with consequences for the optical and transport properties of freshly-emitted soot. Inference of $D_{TEM}$ was much more variable, often related to the spread in the data cloud, resulting in uncertainties stemming from the classifier-primary particle sizing pairing of ~85%.

EDM-WS always predicted larger primary particle sizes, a consequence of grouping together central, clumpy regions of the aggregates. This indicates EDM-WS does not so much predict the primary particle size as a different characteristic of the aggregates, which, when combined with the other primary particle sizing techniques, may also be useful in better representing soot aggregate structure at multiple scales (which may be particularly useful when considering surface growth on these clumpy regions).

Future work will examine other characteristics that can be inferred from the provided segmentations, including the number of primary particles, the fractal parameters, and the degree of primary particle overlap; will consider improving the lower detectability limit attainable by the CNN; and will consider transfer learning as a route to improve robustness when considering different (e.g., aged) aerosols, if the current implementation is found to be insufficient.

## Acknowledgements

This work was financially supported by the Natural Sciences and Engineering Research Council of Canada (FlareNet Research Network; PDF-516743-2018), the Canadian Council for the Arts (Killam Postdoctoral Fellowship), and the IGF promotion plan 20226 N of the DECHEMA Research Foundation, which was funded by the German Federation of Industrial Research Associations within the program for Industrial Corporate Research of the German Federal Ministry for Economic Affairs and Energy, based on an enactment of the German Parliament.

The authors would like to acknowledge Lawrence Zhou for preliminary investigations related to this work.

No carbon dioxide was emitted due to the training of neural networks for this publication, thanks to the use of renewable energy.

## CRediT

Timothy A. Sipkens – Conceptualization, Methodology, Software (analysis, post-processing), Formal analysis, Writing - Original Draft; Visualization

Max Frei – Conceptualization, Methodology, Software (CNN), Writing - Original Draft, Writing - Review & Editing, Visualization

Alberto Baldelli – Writing - Review & Editing

P. Kirchen – Supervision, Funding acquisition

Frank E. Kruis – Supervision, Funding acquisition

Steven N. Rogak – Supervision, Funding acquisition, Writing - Review & Editing

# Supplemental information

**Table S1:** Relevant hardware of the utilized GPU server.

| | |
|---|---|
| **Mainboard** | Supermicro X11DPG-QT |
| **CPU** | 2 × Intel Xeon Gold 5118 |
| **GPU** | 4 × NVIDIA GeForce RTX 2080 Ti |
| **RAM** | 12 × 8 GB DDR4 PC2666 ECC reg. |
| **SSD (OS)** | Micron SSD 5100 PRO 960 GB, SATA |
| **SSD (data)** | Samsung SSD 960 EVO 1 TB, M.2 |

**Table S2:** Relevant software of the utilized GPU server.

| | |
|---|---|
| **OS (host)** | Ubuntu 18.04.3 LTS |
| **Python** | 3.8.5 |
| **PyTorch** | 1.6.0 |